\documentstyle[12pt]{article}
\begin{document}
\begin{center}
{\large\bf A Model for Stars of Interacting Bosons and Fermions}
\vspace{1.5cm}

Cl\' audio M. G. de Sousa
\footnote{ On leave from Departamento de F\'\i sica,
Universidade de  Bras\'\i lia, 70910-900, Bras\'\i lia-DF, Brazil
{\it and } International Centre of
     Condensed Matter Physics,  Universidade de Bras\'\i lia ,
     Caixa Postal 04667, \mbox{70919-970,}
     Bras\'\i lia -- DF, Brazil (e-mail: claudio@iccmp.br)}
 and J. L. Tomazelli
    \footnote{On leave from Instituto de F\'\i sica Te\' orica--UNESP, Rua
Pamplona 145, 01405--900, S\~ ao Paulo--SP, Brazil
(e-mail:tomazell@axp.ift.unesp.br)}
\vspace{0.5cm}

{\small International Centre for Thoretical Physics, P.O. Box 586, 34100}\\
Trieste, Italy
\end{center}
\vspace{1.5cm}

\begin{small}
\begin{center}
{\bf Abstract}
\end{center}
\vspace{0.5cm}

In this paper we introduce a current-current type
interaction term in the Lagrangian density of gravity coupled to
complex scalar fields, in the presence of a degenerated Fermi gas. For
low transferred momenta such a term, which might account for the
interaction among boson and fermion constituents of compact stellar
objects, is subsequently reduced to a quadratic one in the scalar sector.
This procedure enforces the use of a complex radial field counterpart in
the equations of motion. The real and the imaginary components of the
scalar field exhibit different behaviour as the interaction increases. The
results also suggest that the Bose-Fermi system undergoes a BCS-like phase
transition for a suitable choice of the coupling constant.

\end{small}

\newpage

{\large\bf 1.\, Introduction}
\vspace{1.0cm}

In modern approaches to astrophysics and cosmology, concepts
that were initialy restricted to elementary particle physics acquire a
wider meaning, particularly in inflationary models. These models give
predictions for the mass density of the present Universe larger than
that observed, if one presumes that it is close to the critical value.
This suggests that there is a large amount of hidden matter, which has
not been detected so far.

Among the possible candidates for the so called dark matter are boson
stars${}^{[1]}$, which consist of gravitational bound states of scalar
particles. Possibly, such structures were formed through gravitational
collapse in the early Universe and may appear also in the core of
composite objects, whose external envelope are made of standard baryonic
matter.

Since in addition to the bosons there were also fermions in the
primordial gas, we would expect boson-fermion stars to prevail.
This system was studied in detail by Henriques {\it et al} ${}{[2]}$.
In their work it has been shown that the properties of boson-fermion stars
are qualitatively the same, irrespective of the addition of a self-coupling
term for bosons. Nevertheless, it seems that there is no modeling for
explicitly dealing with the interaction between bosons and fermions in such
systems.

In this work we introduce an effective coupling between bosons and
fermions to afford a more realistic description of the system and compare
our results with the ones in the current literature.  In section 2 we
construct the energy-momentum tensor for interacting fermions and bosons
using the Schwarzschild metric. In section 3 we obtain the evolution
equations for the coefficients of the metric and for the fields. In
section 4 we exhibit the results of numerical simulations for the
corresponding dynamical system. Finally, in section 5 we discuss our
results and compare with those obtained without taking the interaction
into account.

\vspace{1.0cm}

{\large\bf 2.\, The Boson-fermion Interaction}
\vspace{1.0cm}

Before introducing the interaction term, we outline the
non-interactive boson-fermion model. We assume the metric to be the
standard Schwarzschild one,

\begin{equation}
     ds^{2}=-B(r)dt^{2}+A(r)dr^{2}+r^{2}d\theta ^{2}+r^{2} \sin ^{2}
            \theta d\varphi ^{2} \,\, .
\end{equation}

Our sign and conventions are the same as those used by Liddle ${}^{[3]}.$
The Lagrangian for the complex scalar field with no explicit interaction is
given by

\begin{equation}
{\cal L}= \frac{R}{16\pi G}
-\partial_{\mu}\Phi ^*\partial ^{\mu}\Phi
-m^{2}\Phi^*\Phi \,\,,
\end{equation}
where
\begin{equation}
\Phi(r,\tau)=\phi(r)e^{-i\omega \tau}
\end{equation}
and for our purposes $\phi$ is a complex scalar field.

For the combined bosons-fermions star we consider that the fermions are
described by a perfect fluid with energy density $\rho $ and pressure $p$
as proposed by Chandrasekhar ${}^{[4]}$:

\begin{eqnarray}
\rho &=&K (\sinh t -t) \,\,,                               \nonumber \\
                                                       \\
   p &=& \frac{K}{3}(\sinh t -8\sinh\frac{t}{2}+3t)\,\, ,  \nonumber
\end{eqnarray}
where $t$ is a parameter, $K=m^{4}_{n}/32\pi^{2}$, and $m_{n}$ is the
fermion mass (to be considered as that of neutrons for illustrative
purposes). The evolution equation for the fermions is given by
an equation of state, namely we have

\begin{equation}
p'=-\frac{1}{2}(\rho +p)\frac{B'}{B} \,\,,
\end{equation}
where the primes stand for derivatives whith respect to $r$.

The corresponding energy-momentum tensor for the bosons and fermions
without interaction reads as
\begin{equation}
T_{\mu \nu}^{(0)}=T_{\mu \nu}^{B}+T_{\mu \nu}^{F} \,\,,
\end{equation}
with
\begin{eqnarray}
T_{\mu \nu}^{B} &=& \partial_{\nu}\Phi ^*\partial_{\mu}\Phi +
\partial_{\mu}\Phi ^*\partial_{\nu}\Phi
-g_{\mu\nu}(\partial_{\lambda}\Phi^* \partial ^{\lambda}\Phi
+m^2{\Phi}^*\Phi ) \,\,,  \\
T_{\mu \nu}^{F} &=& (\rho+p)u_{\mu}u_{\nu}+pg_{\mu \nu}  \,\,,
\end{eqnarray}
where superscripts $B$ and $F$ label bosons and fermions from now on.

At this point we introduce the following interaction term in the Lagrangian
density
\begin{equation}
{\cal L}^{int}=\lambda J_{\mu}(\Phi)j^{\mu}(\psi) \,\,,
\end{equation}
where
\begin{eqnarray}
J_{\mu}(\Phi) &=& i({\Phi}^*\partial_{\mu}\Phi -
\Phi \partial_{\mu}{\Phi}^*) \,\,, \\
j^{\mu}(\psi) &=& \overline{\psi}{\gamma}^{\mu}\psi
\end{eqnarray}
which represent the boson and fermion currents, respectively, while the
$\gamma$'s are
the usual Dirac matrices, which satisfy $\gamma^{0 \dagger}=-\gamma^0$ and
$\gamma^{i \dagger}=\gamma^i\,\,(i=1,2,3)$. This is a typical contact or
current-current interaction between bosons and fermions where the
coupling constant has dimension $[\lambda]=M^{-2}$. We emphasize
that in dealing with a nonrenormalizable interaction term an energy scale
must be introduced when we consider quantum corrections; we find a
similar situation in the pure Einstein-Hilbert gravity.
Note that this point is of no relevance since we are giving a
semiclassical treatment to the problem.

The complete Lagrangian density, including the interaction term, is
invariant under global $U(1)$ gauge transformations. For low transferred
momenta we may replace Eq.(8) by
\begin{equation}
j^{\mu}(\psi)=\overline{\psi}\Gamma^{\mu}\psi \,\,,
\end{equation}
where
\begin{equation}
\Gamma^{0} \equiv iu^0 \,\,\,\,\,,\,\,\,\,\,\Gamma^i \equiv u^i \,\,.
\end{equation}
The four-vector $u^{\mu}=(u^0,u_r,u_{\theta},u_{\varphi})$ is the
four-velocity of the fermion fluid.

The contribution of the interaction term for the energy-momentum tensor
is given by
\begin{equation}
T_{\mu \nu}^{int}=-i\lambda (\Phi^* \partial_{\mu}\Phi -\Phi\partial_{\mu}
                    \Phi^* )\overline{\psi}\Gamma_{\nu}\psi +
           g_{\mu\nu}J_{\lambda}j^{\lambda} \,\,,
\end{equation}
which together with (6) give the total energy-momentum tensor
$$T_{\mu \nu}=T_{\mu \nu}^{0}+T_{\mu \nu}^{int} \,\,.$$

Using (3) and its complex conjugate in (14) we arrive at
\begin{eqnarray}
T_{0}^{0\,\, int} &=& i\frac{\lambda}{\sqrt{A}}( \phi^*\phi '-\phi\phi^*{'})
                  \overline{\psi}\psi \,\,, \\
T_{1}^{1\,\, int} &=& -2\omega\frac{\lambda}{\sqrt{B}}\phi^*
                  \phi \overline{\psi}\psi \,\,.
\end{eqnarray}

Instead of the bilinear operator $\overline{\psi}\psi $ we take the
ground-state of the fermions system  $<\overline{\psi}\psi >_F$ in the
semi-classical approximation. In this limit we replace this quantity
by the average fermion density $\overline{n}_{F}$.

Next, we write the complex scalar fields as
$$\phi=\phi_1+i\phi_2 \,\,\,\,,\,\,\,\, \phi^*=\phi_1-i\phi_2 \,\,,$$
which yields
\begin{eqnarray}
T_{0}^{0\,\, int} &=& \frac{\alpha}{\sqrt{A}}(\phi_2\phi_1 '
                    - \phi_1 \phi_2 ') \,\,, \\
T_{1}^{1\,\, int} &=& -\frac{\omega\alpha}{\sqrt{B}}
                    ({\phi}_{1}^{2}+{\phi}_{2}^{2})\,\,,
\end{eqnarray}
where $\alpha \equiv 2\overline{n}_F \lambda$. The remaining nonvanishing
contributions for the total energy momentum-tensor follows immediately
from (6).

\vspace{1.0cm}

{\large\bf 3.\, Evolution Equations}
\vspace{1.0cm}

Now we are ready to write the evolution equations for our system. With an
appropriate redefinition of the dynamical variables and parameters,
namely
\begin{eqnarray}
                  x &=& mr  \,\,,     \nonumber \\
          \sigma(x) &=& \sqrt{8\pi G}\phi(r) \,\,, \nonumber \\
 \overline{\rho}(t) &=& \frac{4\pi G}{m^2}\rho(t) \,\,,\nonumber \\
    \overline{p}(t) &=&  \frac{4\pi G}{m^2}p(t) \,\,,\nonumber
\end{eqnarray}
$$\overline{\alpha} = \frac{\alpha}{m} \,\,\,,\,\,\,
                  w = \frac{\omega}{m}\,\,,$$
the equations read
\begin{eqnarray}
         A'&=&xA^2\left[ 2\overline{\rho}+\left(
              \frac{w^2}{B}+1\right) \sigma^2+\frac{s^2}{A}
              -\frac{\overline{\alpha}}{\sqrt{A}}(\sigma_2\sigma_1 '
              - \sigma_1 \sigma_2 ') \right] -\frac{A}{x}(A-1) \,\,,  \\
         B'&=&xAB\left[ 2\overline{p}+\left( \frac{w^2}{B}-1\right)
              \sigma^2+\frac{s^2}{A}
              -\frac{w\overline{\alpha}}{\sqrt{B}}\sigma^2 \right]
              +\frac{B}{x}(A-1)  \,\,,  \\
\sigma_1{'}&=&s_1  \,\,,  \\
\sigma_2{'}&=&s_2 \,\,,  \\
     s_1{'}&=&-A\left(\frac{w^2}{B}-1+\frac{\overline{\alpha}w}{\sqrt{B}}
              \right) \sigma_1 +\overline{\alpha}\sqrt{A}s_2+
              \left[ \frac{1}{2}\left( \frac{A'}{A}-\frac{B'}{B}\right)
              -\frac{2}{x}\right] s_1                      \nonumber \\
           & &+\frac{\overline{\alpha}\sqrt{A}}{2}\left(
               \frac{B'}{2B}+\frac{2}{x}\right) \sigma_2 \,\,,  \\
     s_2{'}&=&-A\left( \frac{w^2}{B}
              -1+\frac{\overline{\alpha}w}{\sqrt{B}}\right) \sigma_2
              -\overline{\alpha}\sqrt{A}s_1+
              \left[ \frac{1}{2}\left( \frac{A'}{A}-\frac{B'}{B}\right)
              -\frac{2}{x}\right] s_2                 \nonumber  \\
           & &-\frac{\overline{\alpha}\sqrt{A}}{2}\left(\frac{B'}{2B}
              +\frac{2}{x}\right) \sigma_1 \,\,,   \\
         t'&=&-2\frac{B'}{B}\frac{\sinh t
              -2\sinh\displaystyle{\frac{t}{2}}}{\cosh t
              -4\cosh\displaystyle{\frac{t}{2}}+3}   \,\,,
\end{eqnarray}
where
$$\sigma^2=\sigma_1^2+\sigma_2^2 \,\,\,, \,\,\, s^2=s_1^2+s_2^2 \,\,.$$

These equations form a non-autonomous system of non-linear first-order
differential equations, which cannot be linearized due to the quadratic
terms involved. Equations (19) and (20) are the Einstein equations;
equations (21)--(24) correspond to the Klein-Gordon equation, whereas
(25) gives the evolution of the fermion energy density and pressure.
In the above set of equations the dynamical variables and parameters are
dimensionless. From now on the primes stand for derivatives with respect to
$x$.

It is noteworthy that these equations are invariant under the scale
transformation $$B\longrightarrow \eta B \,\,\,\,,\,\,\,\,
   w\longrightarrow \sqrt{\eta}w \,\,,$$
even after the inclusion of the interaction term. Since the
initial value $B_0$ is undetermined, this permits its redefinition
during numerical calculations, in such a way
to obtain the asymptotic values for the metric coefficients converging to
those of a flat space.

\vspace{1.0cm}
{\large\bf 4.\, Numerical Results}
\vspace{1.0cm}

In this section we present the numerical simulation results of
the above set of equations.
The set of equations were solved by using the fifth-order Runge-Kutta
method; we also use an apropriate ``shooting method'' to infer the value
of $w$, according to an initial value of $B_0$. For further
details on the numerical criteria, the interested reader is referred
to Ruffini and Bonazzola${}^{[5]}$. We require that the metric to be
asymptotically flat and that the scalar fields as well as their
derivatives to vanish at infinity.

Close to the singularity at the origin, which is inherent to Schwarzschild
metric, we see that the last terms in (19) and (20) are dominant. In this
region  $A\sim x/(x-const.)$ and $B\sim (x-const.)/x$, so that $B$ and $B'$
diverges and $A$ becomes oscillating. Hence, for numerical purposes it is
convenient to start with $x$ slightly shifted from the origin.

In Fig.1 we show the results first obtained in Ref.[5] for a typical
boson star, where we note that there is no overlap between $\sigma$ and
$\sigma '$, which characterizes the ground state of the system.
Throughout the simulations the initial values are:
$\sigma_1^0=\sigma_2^0=0.32$, $s_1^0=s_2^0=0$ and $A_0=1$.

In Fig.2 we show the ground-state of a boson-fermion star for $t_0=7$. In
this case $B_0=0.036$ and $w=1.098$.

Fig.3 displays the same curves for $t_0=8.7$ which corresponds
to $B_0=0.01$ and $w=1.327$. Notice that the peak in the curve for $A$
has increased and has been shifted towards the origin, while $\sigma$
and $\sigma '$ approach to zero faster.
Larger values of $t_0$ mean higher fermion energy densities, so that the
fermion contribution to the energy-momentum tensor becomes dominant. In this
sense we would expect curves like those given in Fig.4 to be in agreement
with the pioneering results of Oppenheimer and Volkoff for neutron stars.

Figures 5 and 6 exhibit the results after introducting  the
interaction term, for the choice $t_0=8.7$, and different values of
the dimensionless coupling constant $\overline{\alpha}$. When we switch the
interaction on at small values of $\overline{\alpha}$, $w$ increases, as
shown in table 1. On the other hand $t$ vanishes at a smaller $x$.

If we continue to increase the interaction we observe that, for a certain
value of $\overline{\alpha}$, $w$ suddenly decreases, suggesting that
there is a critical value of $\overline{\alpha}$, in the range
$10^{-3} - 10^{-2}$, in which the system experiences a second-order phase
transition, corresponding to boson-fermion pair formation. This
phenomenum is similar to that observed in the BCS theory; this is not so
surprising since we have started with a quartic interaction term of the
form (9). In Fig.6 (b) and (c) we can observe the splitting of the
real and imaginary parts of the scalar field that also occurs in such
interval. For completeness, we also exhibit the phase space diagrams for
the metric coefficients and for the scalar fields at
$\overline{\alpha}=10^{-2}$ in Fig.7.

\vspace{1.0cm}

\begin{tabular}{|l|l|l|}   \hline
\multicolumn{1}{|c}{$\overline{\alpha}$} &
\multicolumn{1}{|c}{$w$} &
\multicolumn{1}{|c|}{$B_0$} \\
\hline
  $10^{-15}$       & 1.482      & 0.11 \\
  $10^{-9}$        & 1.557      & 0.11 \\
  $10^{-6}$        & 1.565      & 0.011 \\
  $10^{-3}$        & 1.335      & 0.010 \\
  $10^{-2}$        & 1.060      & 0.009 \\
\hline
\end{tabular}

\vspace{0.5cm}

{\small {\bf Table 1:} values of $w$ an $B_0$ for different coupling
constants $\overline{\alpha}$ and $t_0=8.7$}.

\newpage

{\large\bf 5.\, Concluding Remarks}
\vspace{1.0cm}

In this work we studied a model for boson-fermion stars in
interaction, in the region of low frequencies, and observed the behaviour
of the system for increasing values of the coupling constant
$\overline{\alpha}$. At small values of $\overline{\alpha}$ there is no
significant changes in the system; as $\overline{\alpha}$ becomes larger
and larger the ground state energy of the system increases and the fermion
energy density and pressure vanish at smaller distances. As a result, the
fermion star is confined to a smaller region, after which the scalar
fields are still present.

When $\overline{\alpha}$ reaches a critical value, the ground state
energy of the system suddenly decreases, indicating the possible
occurence of a second-order phase transition. In this case, we might have
supplied enough energy to bind bosons and fermions together.

{}From the results outlined above it is possible to compute the mass and the
radius for typical boson-fermions stars; this is the subject of a
forthcoming paper. Another interesting matter, is to develop a
qualitative approach to our evolution equations with Schwarzschild
metric, in order to compare the phase diagrams with our corresponding
numerical results.

A more general treatment of boson-fermion gravitationally bounded
systems should incorporate spin effects by considering the
full Dirac Lagrangian. However, it would be necessary to extend the
system of differential equations from seven to fifteen, which might
demand a great effort.

\newpage

{\bf Acknowlegements}
\vspace{1.0cm}

We are indebted to Professor Ali Awin for carefully reading
the manuscript and to
the International Centre for Theoretical Physics for the hospitality.
CMGS would like to thank CNPq and JLT would like to thank CAPES
for the financial support.

\vspace{1.0cm}

{\large\bf References}
\vspace{1.0cm}

\begin{description}
   \item[1.] Ph. Jetzer, {\em Phys. Rep.\/} {\bf 220}, 4 (1992) 163;

   \item[2.] A. B. Henriques, A. R. Liddle and R. G. Moorhouse, {\em Phys.
             Lett.\/} {\bf B 233} (1989) 99;

   \item[3.] A. R. Liddle and M. S. Madsen, {\em Int. J. Mod. Phys.\/}
            {\bf D 1}, 1 (1992) 101;

   \item[4.] S. Chandrasekhar, {\em Ap. J.\/} {\bf 140},
             2 (1964) 417;

   \item[5.] R. Ruffini and S. Bonazzola, {\em Phys. Rev.\/} {\bf 187}
            (1969) 1767;

   \item[6.] J. R. Oppenheimer and G. M. Volkoff, {\em Phys. Rev.\/}
             {\bf 55} (1939) 374.
\end{description}

\vspace{1.5cm}

{\large\bf Figures Captions}
\vspace{1.5cm}

\begin{description}
\item[Figure 1:] The values of the fields and metric coefficients in a
typical boson star configuration. The corresponding value of $w$ is around
0.826.

\item[Figure 2:] Fields and metric coefficients for a typical
boson-fermion star, with $t_0=7$. In this case, the value of $w$ has
increased to 1.098.

\item[Figure 3:] The same curves as those of Fig.2, for $t_0=8.7$.

\item[Figure 4:] The evolution of $t$ for different initial values.

\item[Figure 5:] Fields and metric coefficients for a boson-fermion star
when we switch on the interaction at $\overline{\alpha}=10^{-9}$.

\item[Figure 6:] The same curves as those shown in the last figure,
for $\overline{\alpha}=10^{-2}$.

\item[Figure 7:] Phase space diagrams (a) for the metric coefficients and
(b) for the scalar fields.
\end{description}
\end{document}